\documentclass[12pt, fleqn, preprint, preprintnumbers, pp, showpacs, showkeys]{revtex4}
\usepackage{epsfig, bm, amsmath, amsfonts}
\usepackage[below]{placeins}
\usepackage{threeparttable}

\def\ltsima{$\; \buildrel < \over \sim \;$}
\def\ltsim{\lower.5ex\hbox{\ltsima}}
\def\gtsima{$\; \buildrel > \over \sim \;$}
\def\gtsim{\lower.5ex\hbox{\gtsima}}

\newcommand{\paren}[1]{\left(#1\right)}
\newcommand{\brak}[1]{\left[#1\right]}

\newcommand{\lparen}[1]{\left(#1\right.}
\newcommand{\rparen}[1]{\left.#1\right)}

\newcommand{\braket}[1]{\left<#1\right>}

\newcommand{\D}{\displaystyle}

\newcommand{\parti}[2]{\frac{\partial#1}{\partial#2}}
\newcommand{\partin}[3]{\frac{\partial^{#3}#1}{\partial#2^{#3}}}

\newcommand{\di}[2]{\frac{d#1}{d#2}}
\newcommand{\din}[3]{\frac{d^{#3}#1}{d#2^{#3}}}
\newcommand{\sech}{\,\mathrm{sech} \,}

\begin{document}

\title{Axisymmetric Nonlinear Waves and Structures in Hall Plasmas}
\author{Tanim Islam}
\affiliation{Lawrence Livermore National Laboratory, P.~O.~Box 808, Livermore, CA 94551-0808}
\email{islam5@llnl.gov}
\date{\today}

\begin{abstract}
In this paper, a general equation for the evolution of an axisymmetric magnetic field in a Hall plasma is derived, with an integral similar to the Grad-Shafranov equation. Special solutions arising from curvature -- whistler drift modes that propagate along the electron drift as a Burger's shock, and nonlinear periodic and soliton-like solutions to the generalized Grad-Shafranov integral -- are analyzed. We derive analytical and numerical solutions in a classical electron-ion Hall plasma, in which electrons and ions are the only species in the plasmas. Results may then be applied to the following low-ionized astrophysical plasmas: in protostellar disks, in which the ions may be coupled to the motion of gases; and in molecular clouds and protostellar jets, in which the much heavier charged dust in a dusty Hall plasma may be collisionally coupled to the gas.
\end{abstract}

\maketitle

\section{Introduction}
Hall physics is relevant in plasmas in which one or more species of a plasma are nonmagnetized, such as in laboratory plasmas on time scales much shorter than the inverse ion cyclotron frequency \cite{Kingsep84, Fruchtman92, Huba94, Gordeev94, Huba95}. Within the past thirty years there have been numerous theoretical, experimental, and numerical studies of Hall effects in plasmas. Experimental, numerical, and theoretical research on Hall plasmas have focused on planar rather than more complicated geometries.

Whistler drift modes propagate as a shock wave via density gradients in a Hall plasma \cite{Kingsep84}. However, the same drift modes, which propagate as shocks, and whistler-like modes with periodicity, may also be seen when considering curvature effects in the absence of density gradients. Whistler-like modes with curvature effects have been shown to be important in magnetic reconnection in the earth's magnetosphere \cite{Shay98, Rudakov02b}, and Hall physics is important in characterizing reconnection \cite{Pritchett01, Rudakov02c}. Linear studies of Hall physics in a low-ionized, high-density protostellar disk\cite{Wardle99, Balbus01} (i.e. cylindrical geometry) have shown that the whistler frequency is also an important parameter in the stability of the rotating disk, and comprehensive numerical studies \cite{Sano02, Sano02b} have borne this out. 

Rudakov and others \cite{Rudakov01, Salimullah99, Rosenberg00} have analyzed whistler and whistler drift modes with planar geometry and inhomogeneities in astrophysical dusty plasmas. This paper continues work done by Rudakov\cite{Rudakov02} and analyzes the nonlinear axisymmetric structure of whistler and whistler drift modes. For simplicity, we construct Hall magnetohydrodynamic (MHD) solutions for an electron-ion plasma, but initially demonstrate their applicability to low-ionized astrophysical plasmas.

The organization of this paper is as follows. In Sec.~(\ref{sec:physics}), we review appropriate dimensional normalizations (length, time, speed, magnetic field) for an electron-ion Hall plasma. In subsequent subsections we describe changes in physics, and corresponding changes in scalings, that expand the applicability of Hall MHD to low-ionized electron-ion-gas plasmas (protostellar disks) and to dusty plasmas (protostellar jets). In Sec.~(\ref{sec:hall_dynamic_equations}) we derive the generalized Grad-Shafranov equation, first introduced in \cite{Rudakov02}, as an interesting class of solutions to an axisymmetric Hall plasma. In Sec.~(\ref{sec:solutions}) we discuss interesting new solutions to the GGS beyond those described in \cite{Rudakov02}. The concluding section, Sec.~(\ref{sec:conclusions}), summarizes this paper's results and illuminates an alternative explanation of what has been though to be magnetic bubbles \cite{Newman92} moving along protostellar jets.

\section{Hall Physics And Length Scales} \label{sec:physics}
In a standard electron-ion Hall plasma, magnetic fields are tied to the electron, rather than fluid, velocity. As a result, unlike in MHD, we may get penetration of the magnetic field into the bulk medium. Furthermore, for this traditional Hall plasma, one gets these scalings
\begin{itemize}
	\item The frequency of Hall phenomena are normalized in units of $\omega_{ci} = Z_i e B_0 / \paren{m_i c}$, the ion cyclotron frequency. $Z_i e$ is the charge of the ion, $B_0$ is the magnetic field strength, $m_i$ is the proton mass, and $c$ is the speed of light.
	
	\item Length scales are in units of $d_i = c/\omega_{pi}$, the ion inertial skin depth, where $\omega_{pi}^2 = 4\pi e^2 n_i/m_i$ is the ion plasma frequency. $n_i$ is the ion number density.
	
	\item Velocities of Hall magnetic structures are in units of $v_{A}$, where $v_{A} = B_0/\sqrt{4\pi n m_i}$ is the Alfv\'{e}n velocity.
	
	\item Magnetic diffusion is given by the following expression,
	\begin{eqnarray}
		&&{\mathcal D}_m = \frac{m_e \nu_{ei} c^2}{4\pi e^2 n_i}. \label{eq:DM}
	\end{eqnarray}
	$\nu_{ei}$ is the electron-proton collision frequency and $m_e$ is the electron mass.
\end{itemize}
These scalings are also summarized in Eq.~(\ref{eq:norm}). Sec.~(\ref{sec:low_ionized_hall_plasmas}) and (\ref{sec:dusty_hall_plasmas}) describe the low-ionized astrophysical plasmas in which Hall MHD phenomena may be observed. Different physics results in different dimensional scalings of Hall MHD phenomena. Most notably, in these low-ionized plasmas, magnetic diffusion is achieved through ion-gas and dust-gas collisionality, rather than electron-ion collisionality (see, e.g., \cite{Harquist97, Balbus01, Rudakov01} for a review).

\subsection{Low Ionized Hall Plasmas} \label{sec:low_ionized_hall_plasmas}
In low ionized plasmas, charged particles make up only a minority of the number and mass density of the total medium. In general the electric field consists of inductive, Hall, Ohmic, and ambipolar components. In order that the low-ionized plasma may be approximated by a resistive Hall MHD electric field (Ohmic and Hall electric fields) with negligible ambipolar electric fields, the ions must remain collisionally coupled to the gas while electrons must remain collisionally uncoupled. Using estimates of those physical regimes in a low-ionized plasma  where either the inductive, Hall, ambipolar, or Ohmic electric fields dominate \cite{Balbus01}, resistive Hall MHD plasma dynamics may occur where $\nu_{eg} \nu_{ig} \gg \omega_{ci} \omega_{ce}$ and $\omega_{ce} \gg \nu_{eg}$. $\nu_{ig}$ and $\nu_{eg}$ are the ion-gas and electron-gas collisional frequencies, and $\omega_{ce}$ is electron cyclotron frequency.

In low-ionized plasmas whose electric fields are resistive Hall MHD in nature, the following substitutions to an electron-ion Hall plasma are made.
\begin{itemize}
	\item We replace $\nu_{ei}$ with $\nu_{eg}$ in Eq. (\ref{eq:DM}).
	
	\item The ions are coupled to the gas, so that the Alfv\'{e}n velocity is calculated over the density of gas particles $v_{Ag} = B_0/\sqrt{4\pi m_g n_g}$, where $m_g$ and $n_g$ are the averaged molecular mass and number density of gas particles, respectively.
	
	\item Since Hall plasma time scales are scaled with respect to the ion cyclotron frequency, the Hall length scale (the largest 
	length scale over which we can observe Hall effects) is given by this relation:
	\begin{eqnarray}
		&&\ell\sim \frac{v_{Ag}}{\omega_{ci}} \approx \frac{c}{\omega_{pi}} \paren{\frac{m_g}{m_i}}^{1/2}\paren{\frac{n_g}
		{n_i}}^{1/2} \label{eq:lengthnorm}
	\end{eqnarray}
\end{itemize}
The collision rates for electrons and ions are given in \cite{Draine83, Chapman56}, where $A$ is the molecular/atomic mass of the principal ion:
\begin{eqnarray}
	&&\begin{aligned}{}
		\nu_{ig} =& n_g\braket{\sigma v}_{ig} \\
		=& 2.6\times10^{-9}\paren{\frac{A}{1\text{amu}}}^{-1/2}\text{ s}^{-1}
	\end{aligned}\label{eq:ioncoupling} \\
	&&\begin{aligned}{}
		\nu_{eg} =& n_g\braket{\sigma v}_{eg} \\
		=& 8.28\times 10^{-10}\paren{\frac{T}{1 K}}^{1/2}\text{ s}^{-1}
	\end{aligned} \label{eq:electroncoupling}
\end{eqnarray}
In a typical protostellar disk, $n_g \sim 10^{12} - 10^{13}$ cm$^{-3}$ \cite{Fromang02}, magnetic fields are of order 10 - 100 mG (see \cite{Myers88} for scaling of magnetic fields in molecular clouds), so that we have a Hall plasma. Most of the ions in low-ionized astrophysical plasmas are alkali ions\cite{Fromang02, Oppenheimer74}, giving an ion mass $m_i \sim 30$. For $T \sim 30 $K, gas density $n_g = 10^{13}$ cm$^{-3}$, assuming a standard cosmic-ray ionization $\zeta = 10^{-17}$ s$^{-1}$\cite{Spitzer68}, ion density $n_i \sim 1$ cm$^{-3}$ \cite{Umebayashi90, Wardle99}. Here $\nu_{ig} \sim 1.4\times 10^5$ cm$^{-3}$. For $B = 100$ mG, $\omega_{ci} = 30$ s$^{-1}$ $\ll \nu_{ig}$. Furthermore, since $\nu_{ei} = 2.6\times 10^4$ s$^{-1}$ and $\omega_{ce} = 1.8\times 10^{6}$ s$^{-1}$, $\nu_{eg}\nu_{ig} \gg \omega_{ci}\omega_{ce}$ and we may completely neglect the ambipolar electric field in the induction equation. The upper threshold for Hall effects $L \sim 5$ AU ($7.5\times 10^{13}$ cm), comparable to the scale of temperature and density gradients in the outer parts of a protostellar disk.

\subsection{Physical Regimes of Dusty Plasmas} \label{sec:dusty_hall_plasmas}
Approximately 1\% of the interstellar medium in our galaxy is composed of dust\cite{Zweibel99}. Rudakov \cite{Rudakov01} showed how to apply the results of electron-ion Hall plasmas to dusty plasmas. It is important to note that although dust may not be a primary charge carrier in a low-ionized astrophysical plasma, their large masses, small charges, and relatively small number densities opens up a length and time scale over which the dust remains unmagnetized (whether through long dust gyroperiod or through collisional coupling with interstellar gas) while the faster and smaller-scale electrons and ions remain fixed to the magnetic field -- hence dusty Hall MHD physics. The following changes to the standard electron-ion Hall MHD scaling must be made,
\begin{eqnarray}
	&&\begin{aligned}{}
		&n_i\to n_d \\
		&e Z_i\to e z_d \\
		&v_i\to v_d \\
		&{\mathcal D}_m\to {\mathcal D}_{m,d} = \frac{\nu_{ig} m_d c^2}{4\pi e^2 n_i\brak{\paren{z_d n_d/n_i}^2 +
		\paren{\nu_{ig}/\omega_{ci}}^2}}.
	\end{aligned} \label{eq:dusttransform}
\end{eqnarray}
$n_d$ and $m_d$ are the number density and mass of dust particles (of a single mass and size), $z_d$ is charge per dust particle, and $v_d$ is the velocity of the dust. In the limit of low temperature plasmas ($T \alt 100$ K), $z_d = -1$. Dusty plasma behaves as a Hall plasma if $n_d/n_i \gg \nu_{ig}/\omega_{ci}$ \cite{Rudakov01}. The dust inertial scale, the lower limit over which we may apply Hall physics, is $\paren{M_d c^2/4\pi e^2 n_d z_d^2}^{1/2} = c/\omega_{pd}$, where $\omega_{pd}$ is the dust plasma frequency.

In the limit that the Lorentz force on a dust particle ($ev_d B/c$) is smaller than the drag force acting on the dust ($a^2 v_{g,th} m_g n_g v_d$, where $a$ is the size of a dust grain):
\begin{eqnarray}
	&&\frac{eB}{m_g c} < a^2 v_g n_g \label{eq:dustcollcrit}
\end{eqnarray}
Then the dust is collisionally coupled to the gas. In this limit, the dusty plasma Hall length scale $c/\omega_{pd}$ increases by a factor $\paren{m_g/m_d}^{1/2}\paren{n_g/n_d}^{1/2}$, similar to the length scale of the low-ionized gas in Eq.~(\ref{eq:lengthnorm}). $m_g$ and $n_g$ are the masses and number densities of gas, respectively.

Dense molecular clouds have a hydrogen gas density of $n_H \sim 10^3 - 10^6$ cm$^{-3}$ (see, e.g. \cite{Harquist97, Zweibel99}). Given a standard cosmic ray ionization $\zeta = 10^{-17}$ s$^{-1}$ \cite{Spitzer68} at temperature $T = 30$ K, this implies an ion density $n_i \sim 10^{-3} - 10^{-2}$ cm$^{-3}$\cite{Umebayashi90, Wardle99}. For these gas densities, the magnetic fields are on the order of $10^{-5} - 10^{-3}$ G \cite{Myers88, Zweibel99}. For a representative molecular cloud density $n_g = 10^3$ cm$^{-3}$ at $T = 10$ K with representative grains of size $a = 0.03$ $\mu$m, we get densities \cite{Umebayashi90} $n_e = n_i = 10^{-3}$ cm$^{-3}$, $n_d = 10^{-8}$ cm$^{-3}$, and magnetic fields $B = 10^{-4}$ G. Assuming densities of dust at 3 gm/cm$^3$, this gives an average dust mass $m_d = 0.5\times 10^{-15}$ gm. Furthermore, the length scale of Hall phenomena is $\paren{M_d c^2/4\pi e^2 n_d z_d^2}^{1/2} = c/\omega_{pd} \sim 1000$ AU. Although the density and temperature model inside protostellar disks is still open to question, we have somewhat better estimates of the density of gas at the edges of protostellar disks and in protostellar jets. We are taking the grain radius of $a = 1\, \mu$m that scatters visible light efficiently. Since we have marginal extinction due to the scattering of light by the dust, $R n_d a^2 \sim 1$ and so $n_d \sim 3\times 10^{-7}$ cm$^{-3}$. The length scale for Hall phenomena calculated over this mass and density is $c/\omega_{pd} \sim 1000$ AU. However, this length scale is smaller than that calculated because the Hall physics is dominated by smaller dust particles. The usual dust size distribution function is $n_d(a) \propto a^{-3.5}$ \cite{Mathis77}, and the length scale $c/\omega_{pd}$ changes as $a^3$.

\section{Hall Dynamic Equations} \label{sec:hall_dynamic_equations}
We consider the electron-ion Hall plasma, using a cylindrical geometry and fluid equation for electrons. The Hall dynamic equations in this form were first written in \cite{Rudakov02}. Here, we begin with the Hall frozen-in equations with magnetic diffusion expression ${\mathcal D}_m$ in Eq. (\ref{eq:DM}), and letting ions remain motionless.
\begin{eqnarray}
	&&\begin{aligned}{}
		&\begin{aligned}{}
		&n\parti{}{t}\paren{\frac{\bf B}{n}} + n{\bf v}_e\cdot\nabla\paren{\frac{\bf B}{n}} = \paren{{\bf B}\cdot\nabla}{\bf v}_e - \nabla{\mathcal D}_m\times\paren{\nabla\times{\bf B}} + {\mathcal D}_m\nabla^2{\bf B}
		\end{aligned} \\
		&\nabla\times{\bf B} = \frac{4\pi}{c}{\bf J} = \frac{4\pi e n}{c}{\bf v}_e \end{aligned} \label{eq:basicHall}
\end{eqnarray}
We consider an axisymmetric magnetic field in cylindrical geometry. Only two variables completely describe the magnetic field: 1) the toroidal magnetic field $B \equiv B_{\phi}$; and 2) the toroidal vector potential $A\equiv A_{\phi}$. This implies an expression for the three-component (poloidal and toroidal magnetic field):
\begin{eqnarray}
    \begin{aligned}{}
    &{\bf B} = B{\bf e}_{\phi} + \nabla\times\paren{A{\bf e}_{\phi}}
    \\
    &B_r = \brak{\nabla\times\paren{A{\bf e}_{\phi}}}_r =
    -\parti{A}{z} \\
    &B_z = \brak{\nabla\times\paren{A{\bf e}_{\phi}}}_z = \frac{1}{r}
    \parti{(rA)}{r} \\
    &B_{\phi} = B
    \end{aligned}
    \label{eq:axisymmB}
\end{eqnarray}
Where $B_r$ and $B_z$ are the radial and vertical components of the
poloidal magnetic field, and $B_{\phi}$ is the toroidal
component.

The electron velocity in terms of the magnetic field is shown here,
\begin{eqnarray}
    \begin{aligned}{}
    &{\bf v}_e = -\frac{c}{4\pi e n}\nabla\times{\bf B} \\
    &v_e^r = -\frac{c}{4\pi e n}\brak{\nabla\times\paren{B{\bf
    e}_{\phi}}}_r = \frac{c}{4\pi e n}\parti{B}{z} \\
    &v_e^z = -\frac{c}{4\pi e n}\brak{\nabla\times\paren{B{\bf
    e}_{\phi}}}_z = -\frac{c}{4\pi e n r}\parti{\paren{rB}}{r} \\
    &\begin{aligned}{}
    v_e^{\phi} =& - \frac{c}{4\pi e n}\brak{\nabla\times
    \paren{\nabla\times\paren{A{\bf e}_{\phi}}}}_{\phi} = \frac{c}{4\pi e n}\paren{\frac{1}{r}\parti{}{r}\paren{
    r\parti{A}{r}} - \frac{A}{r^2} + \partin{A}{z}{2}}
    \end{aligned} \end{aligned} \label{eq:velectron}
\end{eqnarray}
Where $v_e^r$, $v_e^z$, and $v_e^{\phi}$ are the radial, vertical, and
azimuthal components of the electron flow velocity, respectively. With
Eq.~(\ref{eq:velectron}), the continuity equation for the electron
fluid, and Maxwell's equations, the plasma number density is
steady-state $\partial n/\partial t = 0$.

\subsection{Hall Dynamic Equations}
Let ${\mathcal D}_m$ be a constant. Employing Eq.~(\ref{eq:axisymmB}) and Eq.~(\ref{eq:velectron}) we recover these equations,
\begin{eqnarray}
	&&\begin{aligned}{}
		\parti{B}{t} =& -\paren{\parti{\paren{rB}}{z}\parti{}{r} - \parti{\paren{rB}}{r} \parti{}{z}}\frac{B}{nr} + 
		{\mathcal D}_m\paren{\frac{1}{r}\parti{}{r} \paren{r\parti{B}{r}} - \frac{B}{r^2} + \partin{B}{z}{2}} - \\
		&\paren{\parti{\paren{rA}}{z}\parti{}{r} - \parti{\paren{rA}}{r}\parti{}{z}} 
		\frac{1}{nr}\paren{\frac{1}{r}\parti{}{r}\paren{r\parti{A}{r}} - \frac{A}{r^2} + \partin{A}{z}{2}}
	\end{aligned} \label{eq:fullB} \\
	&&\begin{aligned}{}
    	\parti{A}{t} =& -\frac{1}{nr^2}\paren{\parti{\paren{rB}}{z} \parti{}{r} - \parti{\paren{rB}}{r}\parti{}{z}}rA + 
    	{\mathcal D}_m\paren{\frac{1}{r}\parti{}{r}\paren{r\parti{A}{r}} - \frac{A}{r^2} + \partin{A}{z}{2}}
    \end{aligned} \label{eq:fullA}
\end{eqnarray}
if we make these scalings of time, space, magnetic field, density, velocity, and magnetic diffusion ${\mathcal D}_m$ as shown below,
\begin{eqnarray}
	&&\begin{array}{ll}
		t\to t\omega_{ci} & (r,z) \to (r,z)\omega_{pi}/c \\
		B\to B/B_0 & A\to A/A_0 = A\omega_{pi}/\paren{cB_0} \\
		n\to n/n_0 & \omega_{ci} = eB_0/m_i c \\
		\omega_{pi}^2 = 4\pi e^2 n_0/m_i &v\to v/v_{A} \\
		{\mathcal D}_m \to \nu_{ei}/\omega_{ce} &v_{A} = B_0/\sqrt{4\pi n_0 m_i} \\
		\omega_{ce} = eB_0/m_e c &
	\end{array}. \label{eq:norm}
\end{eqnarray}
$B_0$ and $n_0$ are the maximum toroidal magnetic field (or magnetic field strength) and number density, respectively. $\omega_{ce}$, $\omega_{ci}$, $\omega_{pi}$, $v_A$, are the electron cyclotron frequency, ion cyclotron frequency, ion plasma frequency, and Alfv\'{e}n velocity at that point, respecively.

\subsection{Generalized Grad-Shafranov (GGS) Integral}
We show an important integral of the Hall dynamic equations, following  much of the derivation shown in \cite{Rudakov02}. We assume propagating solutions $A\equiv A(r,z-ut)$ and $B\equiv B(r,z-ut)$, where $u$ is a constant normalized velocity, and a radial-dependent number density $n\equiv n(r)$.

Consider force balance equation in some \textit{moving} frame in which the electric ${\bf E}'$ and magnetic ${\bf B}'$ fields are stationary in time. Electron gas pressure and resistivity are ignored.  The electron velocity in a dimensionless form, using Eq.~(\ref{eq:axisymmB}), is given by:
\begin{eqnarray}
    &&\begin{aligned}{} {\bf v}_e =& - \frac{1}{n}\nabla\times{\bf B} = -\frac{1}{nr}\parti{(rB)}{r}{\bf e}_z +
    \frac{1}{n}\parti{B}{z}{\bf e}_r + \frac{1}{n}
    \paren{\frac{1}{r}\parti{}{r}\paren{r\parti{A}{r}} - 
    \frac{A}{r^2} + \partin{A}{z}{2}}{\bf e}_{\phi}
    \end{aligned} \label{eq:velectron0}
\end{eqnarray}
Begin with the dimensionless frozen-in law for the electron fluid
neglecting electron pressure:
\begin{eqnarray}
    &&{\bf E} + {\bf v}_e\times{\bf B} = {\bf 0}
    \label{eq:dimlessfrozen}
\end{eqnarray}
The propagating solution is moving with sufficiently small velocity to
ignore relativisitic effects so that:
\begin{eqnarray}
    &&\begin{aligned}{}
    &{\bf E}' = {\bf E} + {\bf u}\times{\bf B},
    &{\bf B}' = {\bf B}
    \end{aligned} \label{eq:magtransform}
\end{eqnarray}
In this comoving frame the electric field ${\bf E}' = {\bf 0}$, since
electrostatic potentials are neglected. In the laboratory frame the
electric field is given by:
\begin{eqnarray}
    &&{\bf E} = -{\bf u}\times{\bf B}
\end{eqnarray}
As a result, Eq.~(\ref{eq:dimlessfrozen}) can be rewritten as:
\begin{eqnarray}
    &&\paren{{\bf v}_e - {\bf u}}\times{\bf B} = {\bf 0}
    \label{eq:stationarybalance}
\end{eqnarray}
A general case reproduces the generalized Grad-Shafranov (GGS)
integrals with force balance in $r$, $z$, and $\phi$ directions in
Eq. (\ref{eq:stationarybalance}). Sufficiently, force balance in the
radial direction implies:
\begin{eqnarray}
    &&\begin{aligned}{}
    &\paren{\frac{1}{r}\parti{}{r}\paren{r\parti{A}{r}} -
    \frac{A}{r^2} + \partin{A}{z}{2}}B_z + \paren{n u + 
    \frac{1}{r}\parti{(rB)}{r}} = 0 \end{aligned}\label{eq:rbalance}
\end{eqnarray}
Combined with the fact that the solution consists of magnetic surfaces
at constant poloidal flux $\Psi = rA$, and that fact that
$\partial H/\partial r = dH/d\Psi\times \partial\Psi/\partial r$,
imply the GGS integrals. 
\begin{eqnarray}
    &&\begin{aligned}{}
    &r\parti{}{r}\paren{\frac{1}{r}\parti{\Psi}{r}} +
    \partin{\Psi}{z}{2} = -\di{H}{\Psi}\paren{H(\Psi) -
    u\int_0^r n\paren{r'}r'\,dr'} \end{aligned}\label{eq:GS} \\
    &&rB = H\paren{\Psi} - u\int_0^r n\paren{r'}r'\,dr' \label{eq:int1}    
\end{eqnarray}
The first current term $H(\Psi)$ arises in the normal Grad-Shafranov equation. The additional term in the GGS arises due to the electric field (see Eq.~[\ref{eq:stationarybalance}]) acting on the moving structure. This appears as an additional poloidal current, $-u\int_0^r n\paren{r'}r'\,dr'$, in the frame of the propagating solution, where the motionless ions are seen moving with velocity $-u{\bf e}_z$. This ``current'' can appear only where the electrons are moving. Furthermore, radial electric force or ${\bf J}\times{\bf B}$ force due to this ``current'' results in a pressure term $-u\int_0^r n\paren{r'}r'\,dr'\times dH/d\Psi$ in the GGS equation. It must be emphasized that in MHD the magnetic field is frozen into the plasma and cannot move relative to the ions, so there is no additional force. The GGS equations were first found in \cite{Rudakov02} as exact solutions of the dynamic equations for the Hall magnetic field in an axisymmetric geometry, Eq.~(\ref{eq:fullB}) and (\ref{eq:fullA}).

\section{Axisymmetric Hall MHD Solutions} \label{sec:solutions}
Here, we consider solutions of the GGS, that propagate down a cylindrical column. We have three sets of analytic solutions. Sec.~(\ref{subsec:nonlinshock}) details the resistive nonlinear shock whose structure is described by a Burgers equation. Sec.~(\ref{subsec:periodic}) describes a periodic solution to the GGS equation with linear current term $H(\Psi) \propto \Psi$. Sec.~(\ref{subsec:soliton}) numerically estimates, with justification on the basis of the Chandrasekhar-Fermi theorem\cite{ChandrasekharFermi53}, nonlinear localized bubble solutions to the GGS integral. According to this theorem, a localized MHD equilibrium plasma magnetic structure could exist only in the presence of an external magnetic field or with fixed metal wall boundaries.

\subsection{Resistive Nonlinear Shock} \label{subsec:nonlinshock}
\begin{figure}[ht]
    \epsfig{file = 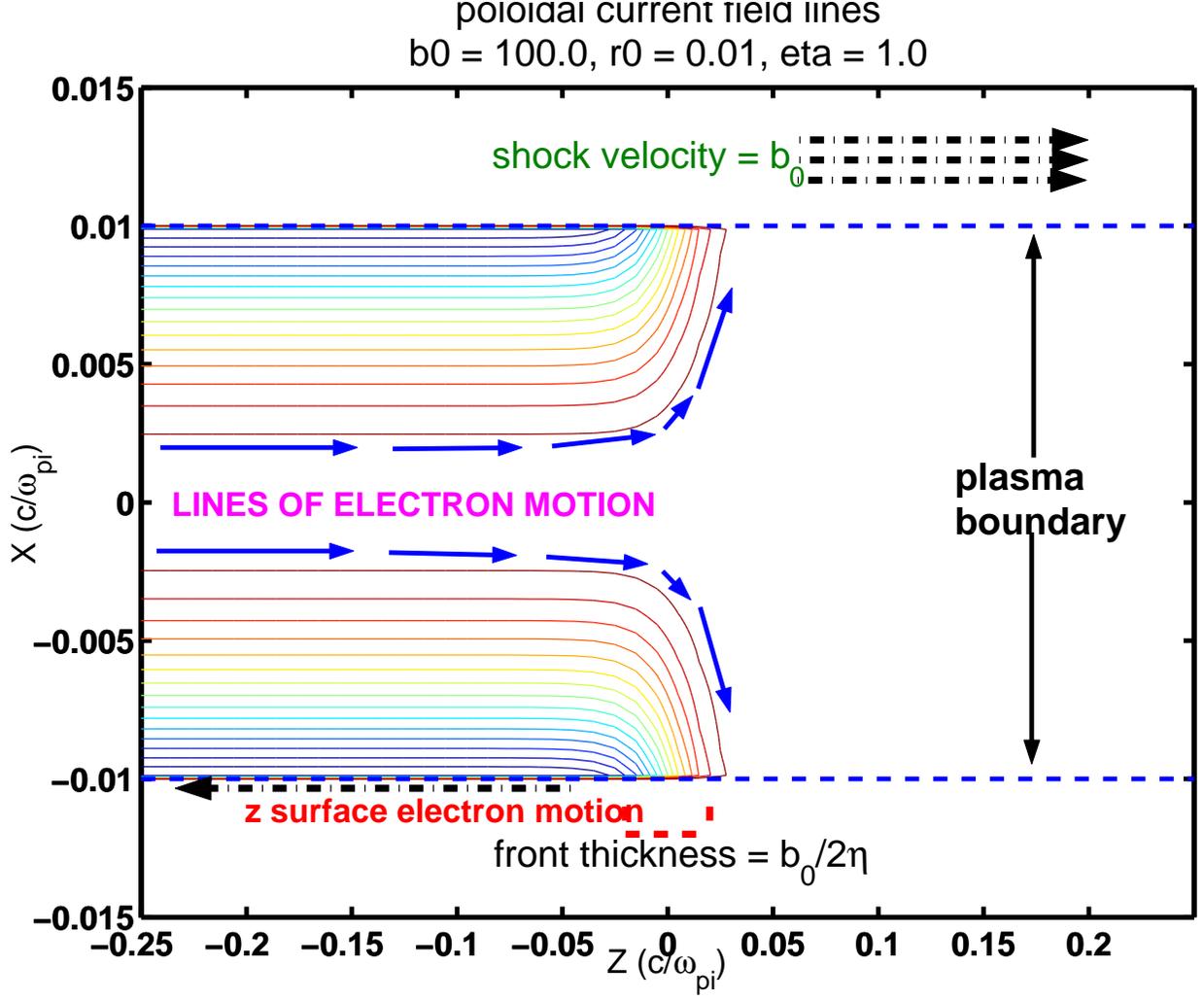, width = \linewidth}
    \caption{Poloidal current lines within the Burgers-like nonlinear
    solution. Here, electrons must return across the surface so, we
    see that a current sheet develops on the Hall plasma column, that
    distributes current back into the (attached) current generator.}
    \label{fig:burgerscurrlines}
\end{figure}

Again, assume a Hall plasma column, so that the magnetic fields are given by Eq.~(\ref{eq:fullA}) and (\ref{eq:fullB}), with a constant-density plasma within the column $r < R$ and vacuum outside. Furthermore, let us assume only the toroidal field $B$ exists with profile $B(r,z,t) = B(z,t)r/R$ and $B(-\infty, t) = -B_0$, where $B_0 > 0$. Thus, we get the Burger's equation in normalized coordinates:
\begin{eqnarray}
    &&\parti{B}{t} = \frac{2B}{R}\parti{B}{z} + {\mathcal
    D}_m\partin{B}{z}{2} \label{eq:case2}
\end{eqnarray}
With this solution, in normalized coordinates:
\begin{eqnarray}
    &&\begin{aligned}{} 
    &B(r,z,t) = 
    \begin{cases} \D\frac{B_0 r}{2R}
    \paren{\tanh\paren{\frac{B_0\paren{z - B_0 t/R}}{2R{\mathcal D}_m}} -
    1} & r < R \\ 0 & r > R \end{cases} \end{aligned}
    \label{eq:burgershock}
\end{eqnarray}
In dimensional values the shock velocity is given by $u = cB_0/\paren{4\pi e n R}$ and shock thickness $\delta = 2\nu_{ei}/\omega_{ce} R$. Note that the shock is moving in the positive $z$ direction in our coordinate system. This asymmetry is typical for Hall phenomena. Furthermore, one may construct analytic solutions for any arbitrary initial configuration $B(z,0)$ (see e.g. \cite{Vainshtein00, Gordeev94} for the Hopf-Cole transformation that transforms Burger's equation into a heat diffusion equation). A plot of the toroidal magnetic structure of this resistive case is shown in figure \ref{fig:burgerscurrlines}, with $u = 100$ and $b_0 = B_0/R = 100$.

\subsection{Periodic GGS Solution} \label{subsec:periodic}
Here, we consider solutions to the GGS equation (see
Eq.~[\ref{eq:int1}] and [\ref{eq:GS}]) with linear current terms,
$H(\Psi) = k\Psi$. Note that this solution arises from a particular
choice of current term, and does not arise from a linearization of the
dynamic equations (Eq.~[\ref{eq:fullA}] and
[\ref{eq:fullB}]). Consider the density $n = 1$ a constant, and as
before $u > 0$ is also a constant. Then the GGS integrals reduce to
these relations inside the plasma column of radius $R$:
\begin{eqnarray}
    &&r\parti{}{r}\paren{\frac{1}{r}\parti{\Psi}{r}} +
    \partin{\Psi}{z}{2} = -k^2\Psi + \frac{1}{2}uk r^2
    \label{eq:GSlin1} \\
    &&B = k A - \frac{1}{2}u r^2 \label{eq:GSlin2}
\end{eqnarray}
Outside the plasma column we have a vacuum, so that there
are no currents and no magnetic field in the frame of electron
MHD. Electron MHD implies that the magnetic field is carried only by
the electrons.

We consider a geometry in which there is a given
radial variation in the three-component magnetic field, such that
there is only a toroidal current sheet. This implies that $B \to 0$ at
the boundary of the plasma column. We look for solutions that go as
$\sim \cos \kappa\paren{z - ut}$ in magnetic field.
\begin{eqnarray}
    &&\begin{aligned}{}
    &\Psi = \begin{cases} \alpha r
    J_1\paren{r\sqrt{k^2 -
    \kappa^2}}\cos \kappa\paren{z-ut} + \frac{ur^2}{\kappa} & r < R \\
    \frac{uR^2}{k} & r > R \end{cases} \\
    &B = \begin{cases}
    \kappa\alpha J_1\paren{r\sqrt{\kappa^2 - k^2}}\cos k\paren{z-ut} &
    r < R \\ 0 & r > R \end{cases} \end{aligned} \label{eq:case3}
\end{eqnarray}
Where $J_n$ are Bessel functions, $\kappa^2 > k^2$, and the condition that
$J_1\paren{R\sqrt{\kappa^2 - k^2}} = 0$. The toroidal surface current
on the plasma column is given by:
\begin{eqnarray}
    &&\begin{aligned}{}
    &I_{\phi}\paren{R,z-ut} = \alpha\sqrt{\kappa^2 - k^2}\lparen{
    J_0\paren{R\sqrt{\kappa^2 - k^2}}} - \\
    &\rparen{J_2\paren{R\sqrt{\kappa^2 -
    k^2}}}\cos k\paren{z-ut} + \frac{2u}{\kappa}
    \end{aligned} \label{eq:torcurr}
\end{eqnarray}

\begin{figure}[!ht]
    \epsfig{file = 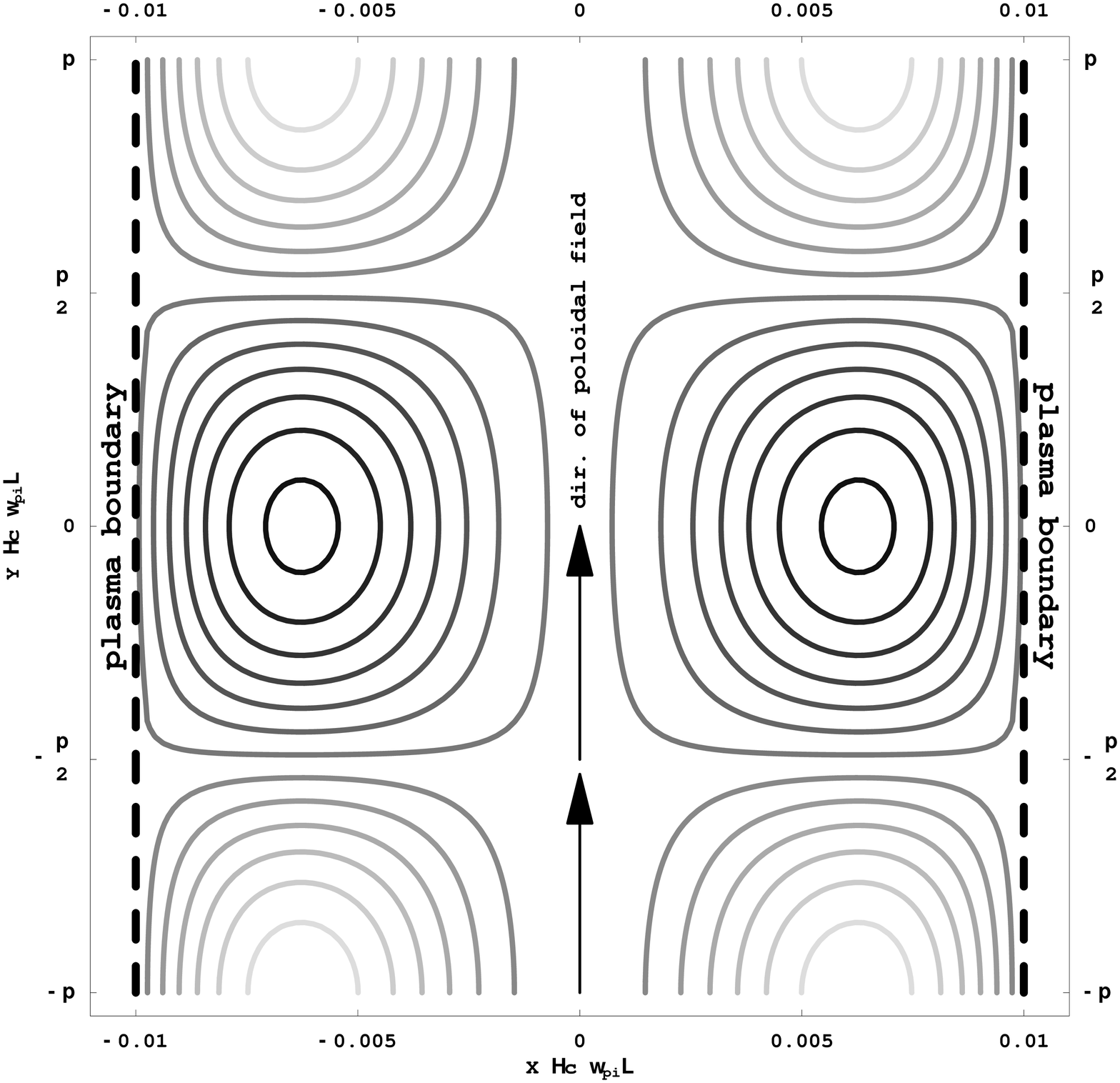, width = \linewidth}
    \caption{Poloidal magnetic flux $\Psi$ (and hence, poloidal magnetic field lines) for the GGS solution. Note that since the poloidal flux within the vacuum is constant, no poloidal magnetic fields exist within the vacuum. Note that the discontinuity in $B_z$ results in $\phi$ current sheet along the Hall column surface. This magnetic configuration appears similar to the compact torus, where the structure is defined by the boundary conditions -- i.e. metal walls.} \label{fig:GSpolsoln}
\end{figure}

The solution shown in figures \ref{fig:GSpolsoln}, with these parameters:
\begin{eqnarray*}
    &&\begin{matrix} \eta = 0.0 & \alpha = 1.0 \\
    \kappa = 1.0 & u = 100.0 \end{matrix}
\end{eqnarray*}
And $\kappa$ given by the first zero of $J_1(x)$:
\begin{eqnarray*}
    &&k = \sqrt{\kappa^2 + \paren{\frac{3.83171}{R}}^2}
\end{eqnarray*}

\subsection{Localized Hall Bubble Solutions} \label{subsec:soliton}
Here, we consider solutions to the GGS equation that describe the evolution of possible 2D nonlinear whistler-like waves and structures. Consider solutions with normalized density $n = 1$, propagating with velocity $u{\bf e}_z$, and the currents and toroidal magnetic fields in the GGS integral:
\begin{eqnarray}
    &&\begin{aligned}{}
    &H\paren{\Psi} = \alpha\Psi^2  \\
    &rB = H\paren{\Psi} - \frac{1}{2}ur^2 = \alpha \Psi^2 -
    \frac{1}{2}ur^2 \end{aligned}\label{eq:currterm}
\end{eqnarray}
In terms of the toroidal vector potential $A$ and the toroidal magnetic
field $B$ the nonlinear soliton equation is given by,
\begin{eqnarray}
    &&\begin{aligned}{}
    &\partin{A}{r}{2} + \frac{1}{r}\parti{A}{r} - \frac{A}{r^2} -
    ur^2 A + 2r^2\alpha A^3 = 0 \\
    &B = \alpha r A^2 - \frac{1}{2} ur \end{aligned} \label{eq:soliton}
\end{eqnarray}
One can easily show that if we let $A_0(r,z)$ be the solution to the differential equation, $u = 1, \alpha = 1$:
\begin{eqnarray}
    &&\partin{A_0}{r}{2} + \frac{1}{r}\parti{A_0}{r} - \frac{A_0}{r^2}
    - r^2 A_0 + 2r^2 A_0^3 = 0 \label{eq:soliton0}
\end{eqnarray}
Then the scaling for the toroidal vector potential $A\paren{r,z;u,\alpha}$ in terms of $A_0(r,z)$ and the form of the toroidal magnetic field from Eq.~(\ref{eq:soliton}) are given by:
\begin{eqnarray}
    &&A\paren{r,z;\mu,\alpha} = \paren{\frac{\alpha}{u}}^{1/2} A_0\paren{u^{1/4}r, u^{1/4}z} \label{eq:Ascaling} \\
    &&B\paren{r,z;u,\alpha} = \paren{\frac{\alpha}{u}}^{3/2}r A_0\paren{u^{1/4}r, u^{1/4}z}^3 - \frac{1}{2}u r \label{eq:Bscaling}
\end{eqnarray}
Consider a volume to be a cylinder. The soliton structure consists of a three-component magnetic field (a poloidal field and a toroidal field) that is localized about the origin with a span much smaller than the dimensions of the cylindrical volume; thus, $B_r \to 0$, $B_z \to 0$, and $B_{\phi} \to -\frac{1}{2}ur$ at the boundaries of the cylinder. Employing the Chandrasekhar-Fermi theorem -- using ${\bf r}\cdot\paren{{\bf J}\times{\bf B}}$ force balance within the cylindrical volume -- one can show,
\begin{eqnarray}
	&&\begin{aligned}{} 
		&\int_{r = 0}^R\int_{z = -L}^L rB_z^2\,dr\,dz = \\
		&\int_{r = 0}^R \int_{z = -L}^L un(r)r H\paren{\Psi}\,dr\,dz
	\end{aligned} \label{eq:CFbalance}
\end{eqnarray}
Furthermore, since $B_z = -\partial A/\partial r$, a localized solution (where the integral on the left hand side is finite) is possible provided that the right hand side $\ge 0$. To guarantee a localized solution we require that $H(\Psi) \ge 0$. The case analyzed in this section is $H\paren{\Psi} = \Psi^2 \ge 0$.

Note that the form of Eq.~(\ref{eq:soliton}) or (\ref{eq:soliton0}) is as if we consider a nonlinear 3D Schr\"{o}dinger equation with the $m = 1$ azimuthal modes (i.e. solutions $\sim e^{i\phi}$). Here, to estimate the form of the soliton, we use the same variational method and trial functions, described below, for 3D nonlinear optical solitons \cite{Desyatnikov00, Yakimenko05}. First, we derive the Lagrangian that describes the GGS equation with nonlinear current. Then we look for localized solutions with different test functions. One reasonable approximation for the test function in cylindrical coordinates is $A(r,z) = U(r)\sech\mu z$.  A one-dimensional differential equation in $r$ is derived by averaging that Lagrangian with respect to $z$ and searching for a localized one-node solution, presumed to be the main lowest-energy stable solution, for a given $\mu$. We solve for that $\mu$ that minimizes the integral of the averaged Lagrangian over $r$. Another test function is in spherical coordinates, $A(\rho,\theta) = U(\rho)\sin\theta$, where $\rho = \sqrt{r^2 + z^2}$ and $\cos\theta = z/\rho$. Again a one-dimensional equation in $\rho$, and its localized single-node lowest energy solution, is derived by averaging and minimizing the Lagrangian of the GGS. A more detailed description of the estimation method for both trial functions is provided in the Appendix.

\begin{figure}[!ht]
	\epsfig{file = 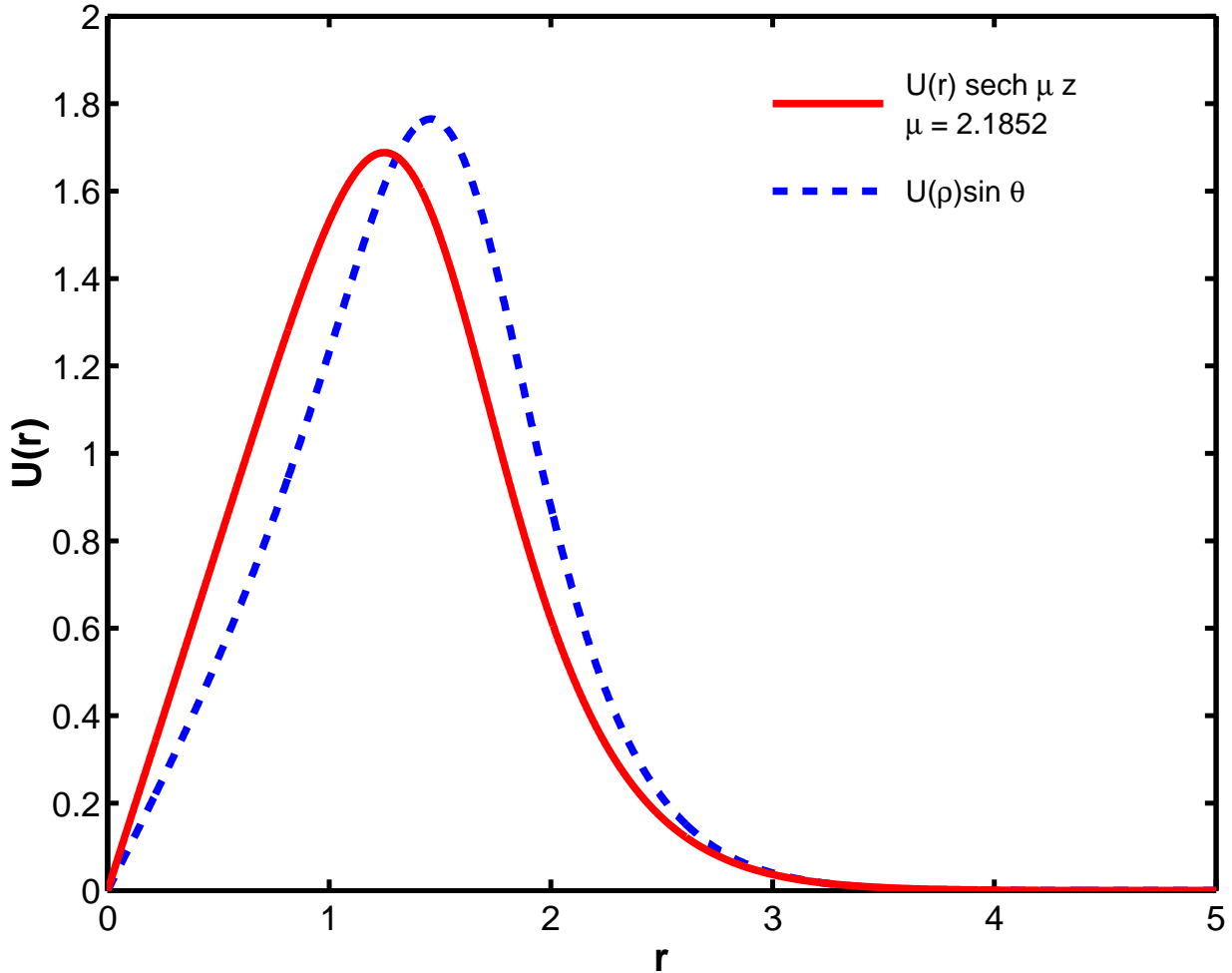, width = \linewidth}
	\caption{Plot of the radial distribution with the given ansatzes shown in the figure. For the cylindrical ansatzes, the $z$ width parameter $\mu$ is calculated to four decimal places.} \label{fig:1drelations}
\end{figure}

\begin{figure}[!ht]
	\epsfig{file = 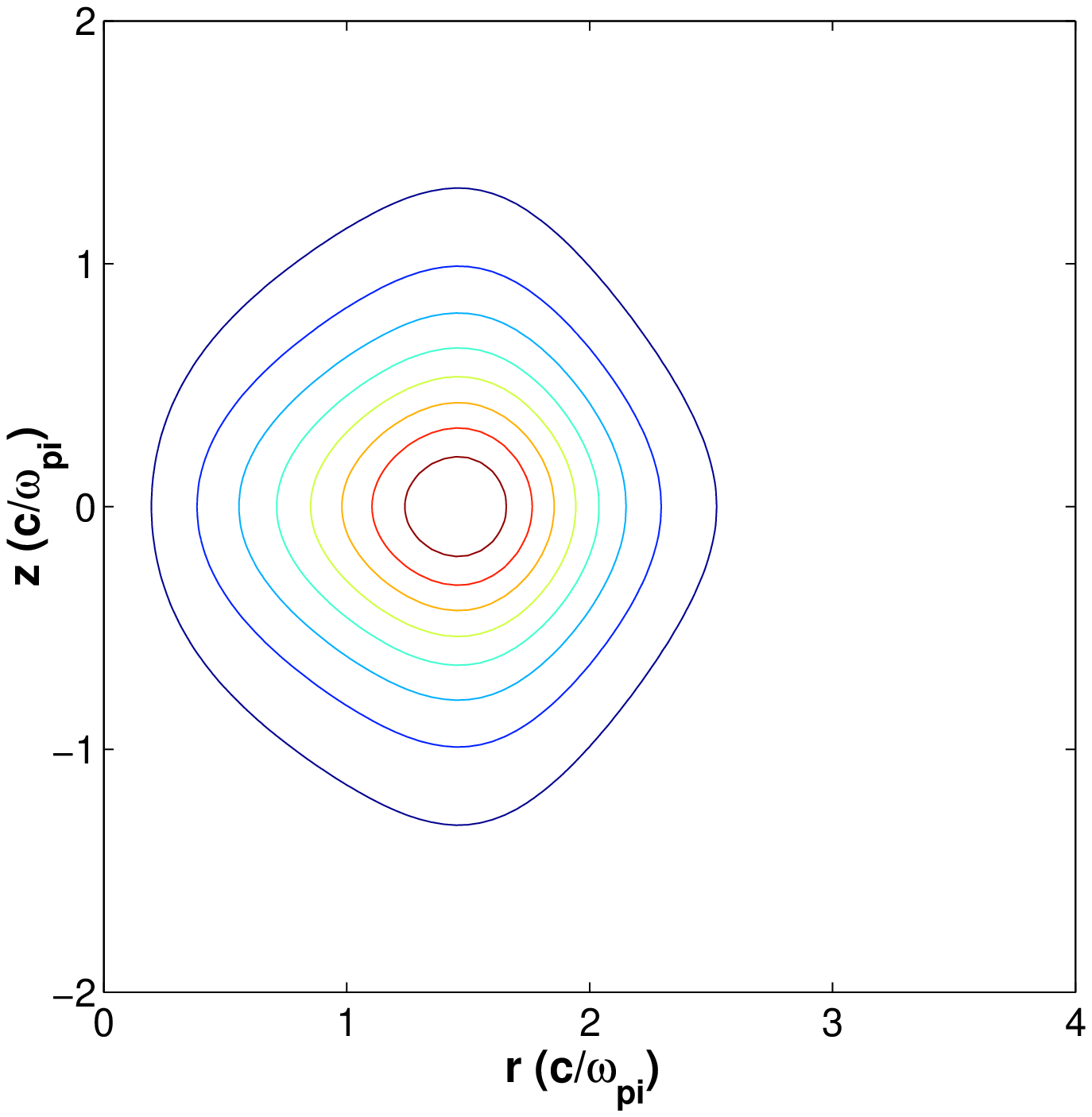, width = 0.48\linewidth}\hfill
	\epsfig{file = 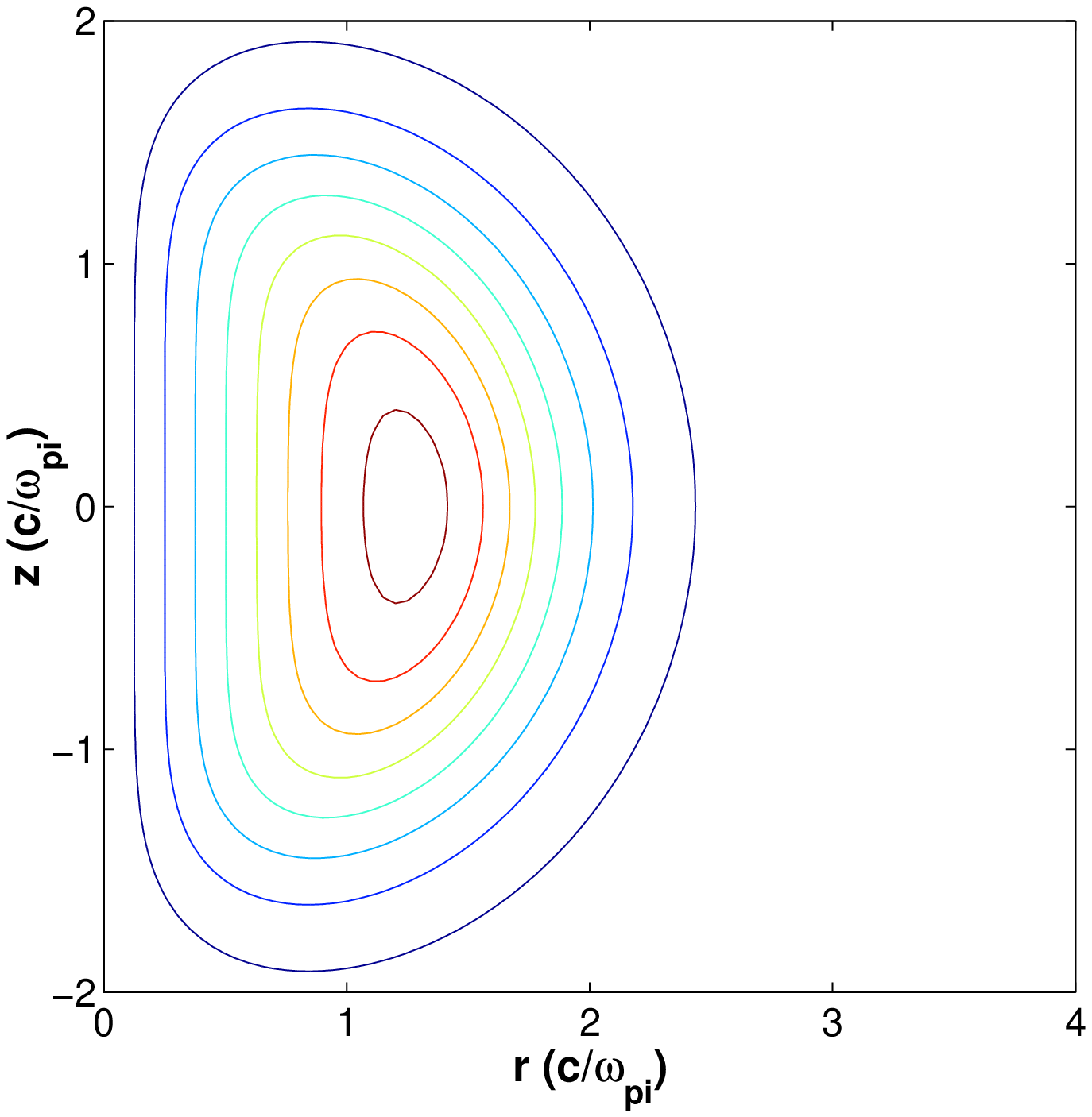, width = 0.48\linewidth}
	\caption{Structure of the $U(r)\text{sech}\,\mu z$ ansatz (left) compared with the spherical $U(\rho)\sin\theta$ ansatz (right). Although the finer features are obviously different, the width in the radial direction are roughly similar.} \label{fig:contours}
\end{figure}

Plots of the dependence $U(r)$, the cylindrical ansatz, and $U(\rho)$, the spherical ansatz, are shown in Fig.~\ref{fig:1drelations}. Contours detailing the shapes of the toroidal vector potential $A(r,z)$ for both trial functions are shown in Fig.~\ref{fig:contours}. The cylindrical ansatz $U(r)\sech\mu z$ and the spherical ansatz $U(\rho)\sin\theta$ have approximate shapes in the radial coordinate. As in \cite{Desyatnikov00}, there is a substantial differences in the shape of the cylindrical and spherical trial functions of the soliton. The shape of the cylindrical and spherical trial functions look similar to their corresponding solutions of the optical solitons found in \cite{Desyatnikov00}. Both trial functions also satisfy the scaling $A\paren{r,z;\mu,\alpha} = \paren{\alpha/u}^{1/2} A_0\paren{u^{1/4}r, u^{1/4}z}$.

\section{Discussion and Conclusions} \label{sec:conclusions}
Here we have explored possible structures that arise in a Hall plasma due to curvature rather than, say, density gradients. We have provided a general structure to continue further analysis of the dynamic Hall modes in Eq.~(\ref{eq:fullB}) and (\ref{eq:fullA}) as well as the generalized Grad-Shafranov (GGS) integrals in Eq. (\ref{eq:GS}) and (\ref{eq:int1}) and propagating solutions with constant velocities in the $z$ direction.

Well-understood Hall plasma phenomena such as whistler drift modes were shown to exist in an homogenous cylindrical plasma column. Furthermore, an exact solution for the whistler drift mode, the Burgers nonlinear shock wave, was discussed in subsection \ref{subsec:nonlinshock}. Periodic solutions were shown in subsection \ref{subsec:periodic} from a simple analysis of the allowable solution for linear current in the GGS integrals. We find that the solution is constrained by the radial vacuum boundary of a Hall plasma column, because in Hall physics the magnetic field is carried solely by the electrons. Further work may involve a search for transitional solutions -- those solutions intermediate between the periodic Grad-Shafranov solution and the collisional nonlinear shock. Possible solutions may be those with oscillatory substructure in the toroidal magnetic field at the shock with corresponding poloidal magnetic field in the case of small resistivity.

We determined that there is a congruence between the GGS and the nonlinear Schr\"{o}dinger equation with $m = 1$ azimuthal modes. We apply the variational method, which was developed for optical fiber solitons, for our problem. We estimate soliton solutions in subsection \ref{subsec:soliton} for a given nonlinear current $H(\Psi) = \alpha\Psi^2$ and constant number density $n$. The Chandrasekhar-Fermi theorem implies that such a nonlinear current may allow for localized soliton solutions. Such a current distribution also allows for a simple scaling -- that is, solutions with normalized propagation velocity $u \ne 1$ and current scalings $\alpha \ne 1$, and $u > 1, \alpha > 1$ have vector potential $A\paren{r,z;u,\alpha} = \paren{\alpha/u}^{1/2}A_0\paren{u^{1/4}r,u^{1/4}z}$. Further research may apply this suggested method to other localized soliton-like solutions, as well as illuminate relationships between the magnetic fields and vertical velocities in propagating localized Hall structures. The Hall dynamic equations in Eq. (\ref{eq:fullB}) and (\ref{eq:fullA}) may be useful in analyzing the stability of the soliton solutions estimated here.

We have also included some space plasmas in which the Hall regime may be important. In Earth's lower ionosphere, whistler drift modes due to density gradients (at the equator) or curvature (at the magnetic poles) may exist. In astrophysical plasmas, such as molecular clouds, protostellar disks, or protostellar jets, Hall phenomena may exist due to the presence of charged dust.

Axisymmetric Hall phenomena may play an important role in the circumstellar disks and jets pictured by the Hubble Space Telescope around young stellar objects. The typical circumstellar disk has a radius of a few hundred AUs -- a possible length scale in Hall phenomena. Hall shocks and other magnetic structures also transport of currents and magnetic fields through matter without compression or net mass motion. A comparison of the radio frequency Doppler studies of the bipolar jets around HH30 show that the bulk velocity of the matter is five times slower than the velocities of the observed jet substructures  \cite{Lada96}. Furthermore, radio polarization measurements of the jets emanating from the disks of GM Aurigae and DG Tauri imply largely toroidal magnetic fields\cite{Tamura99}, which may be explainable as a nonlinear shock wave and structures rather than the traditional model of a supersonic fluid or MHD shock.

Other explanations for the fast-moving, time-resolved substructures in the jets of T Tauri stars involve ``bullets'' of plasma that emanate from the disk \cite{Newman92}. Furthermore, a variety of MHD jet models are given in the literature (see, e.g. \cite{Newman92, Lovelace92, Ustyugova00}). Astrophysical MHD experiments of pressure-confined toroidal magnetic towers (see, e.g., \cite{Ciardi06, Ciardi09}) may explain features of these jets as specific MHD phenomena: higher intensity material within the jet as magnetic bubbles, and longer-timescale intermittency because kink and other azimuthal MHD instabilities collapse the jet structure. Here, due to the obvious presence of dust, we imply an alternative model, magnetic penetration into a Hall plasma, that may explain some of these observed structures. A companion astrophysical paper will expand upon the illuminating relationship between net Hall magnetic bubble velocity, radial dimension, and magnetic field strength to explain the migration of observed ``bullets'' seen in protostellar jets.

\begin{acknowledgments}
The author would like to thank L.~Rudakov for his invaluable advice with regards to these new classes of curvature-based solution in Hall plasmas, and especially for introducing him to the generalized Grad-Shafranov equation.
\end{acknowledgments}
\bibliography{hall}

\begin{thebibliography}{38}
\expandafter\ifx\csname natexlab\endcsname\relax\def\natexlab#1{#1}\fi
\expandafter\ifx\csname bibnamefont\endcsname\relax
  \def\bibnamefont#1{#1}\fi
\expandafter\ifx\csname bibfnamefont\endcsname\relax
  \def\bibfnamefont#1{#1}\fi
\expandafter\ifx\csname citenamefont\endcsname\relax
  \def\citenamefont#1{#1}\fi
\expandafter\ifx\csname url\endcsname\relax
  \def\url#1{\texttt{#1}}\fi
\expandafter\ifx\csname urlprefix\endcsname\relax\def\urlprefix{URL }\fi
\providecommand{\bibinfo}[2]{#2}
\providecommand{\eprint}[2][]{\url{#2}}

\bibitem[{\citenamefont{Kingsep et~al.}(1984)\citenamefont{Kingsep, Mokhov, and
  Chukbar}}]{Kingsep84}
\bibinfo{author}{\bibfnamefont{A.~S.} \bibnamefont{Kingsep}},
  \bibinfo{author}{\bibfnamefont{Y.~V.} \bibnamefont{Mokhov}},
  \bibnamefont{and} \bibinfo{author}{\bibfnamefont{K.~V.}
  \bibnamefont{Chukbar}}, \bibinfo{journal}{Fiz. Plazmy}
  \textbf{\bibinfo{volume}{10}}, \bibinfo{pages}{854} (\bibinfo{year}{1984}).

\bibitem[{\citenamefont{Fruchtman and Rudakov}(1992)}]{Fruchtman92}
\bibinfo{author}{\bibfnamefont{A.}~\bibnamefont{Fruchtman}} \bibnamefont{and}
  \bibinfo{author}{\bibfnamefont{L.}~\bibnamefont{Rudakov}},
  \bibinfo{journal}{Phys. Rev. Lett.} \textbf{\bibinfo{volume}{69}},
  \bibinfo{pages}{2070} (\bibinfo{year}{1992}).

\bibitem[{\citenamefont{Huba et~al.}(1994)\citenamefont{Huba, Grossman, and
  Ottinger}}]{Huba94}
\bibinfo{author}{\bibfnamefont{J.~D.} \bibnamefont{Huba}},
  \bibinfo{author}{\bibfnamefont{I.~M.} \bibnamefont{Grossman}},
  \bibnamefont{and} \bibinfo{author}{\bibfnamefont{P.~F.}
  \bibnamefont{Ottinger}}, \bibinfo{journal}{Phys. Plasmas}
  \textbf{\bibinfo{volume}{1}}, \bibinfo{pages}{3444} (\bibinfo{year}{1994}).

\bibitem[{\citenamefont{Gordeev et~al.}(1994)\citenamefont{Gordeev, Kingsep,
  and Rudakov}}]{Gordeev94}
\bibinfo{author}{\bibfnamefont{A.~V.} \bibnamefont{Gordeev}},
  \bibinfo{author}{\bibfnamefont{A.~S.} \bibnamefont{Kingsep}},
  \bibnamefont{and} \bibinfo{author}{\bibfnamefont{L.~I.}
  \bibnamefont{Rudakov}}, \bibinfo{journal}{Physics Reports}
  \textbf{\bibinfo{volume}{243}}, \bibinfo{pages}{214} (\bibinfo{year}{1994}).

\bibitem[{\citenamefont{Huba}(1995)}]{Huba95}
\bibinfo{author}{\bibfnamefont{J.~D.} \bibnamefont{Huba}},
  \bibinfo{journal}{Phys. Plasmas} \textbf{\bibinfo{volume}{2}},
  \bibinfo{pages}{2504} (\bibinfo{year}{1995}).

\bibitem[{\citenamefont{Shay et~al.}(1998)\citenamefont{Shay, Drake, Denton,
  and Biskamp}}]{Shay98}
\bibinfo{author}{\bibfnamefont{M.~A.} \bibnamefont{Shay}},
  \bibinfo{author}{\bibfnamefont{J.~F.} \bibnamefont{Drake}},
  \bibinfo{author}{\bibfnamefont{R.~E.} \bibnamefont{Denton}},
  \bibnamefont{and} \bibinfo{author}{\bibfnamefont{D.~J.}
  \bibnamefont{Biskamp}}, \bibinfo{journal}{J. Geophys. Res.}
  \textbf{\bibinfo{volume}{103}}, \bibinfo{pages}{9165} (\bibinfo{year}{1998}).

\bibitem[{\citenamefont{Rudakov and Huba}(2002)}]{Rudakov02b}
\bibinfo{author}{\bibfnamefont{L.~I.} \bibnamefont{Rudakov}} \bibnamefont{and}
  \bibinfo{author}{\bibfnamefont{J.~D.} \bibnamefont{Huba}},
  \bibinfo{journal}{Phys. Rev. Lett.} \textbf{\bibinfo{volume}{89}},
  \bibinfo{pages}{95002} (\bibinfo{year}{2002}).

\bibitem[{\citenamefont{Pritchett}(2001)}]{Pritchett01}
\bibinfo{author}{\bibfnamefont{P.~L.} \bibnamefont{Pritchett}},
  \bibinfo{journal}{J. Geophys. Res.} \textbf{\bibinfo{volume}{106}},
  \bibinfo{pages}{3783} (\bibinfo{year}{2001}).

\bibitem[{\citenamefont{Huba and Rudakov}(2002)}]{Rudakov02c}
\bibinfo{author}{\bibfnamefont{J.~D.} \bibnamefont{Huba}} \bibnamefont{and}
  \bibinfo{author}{\bibfnamefont{L.~I.} \bibnamefont{Rudakov}},
  \bibinfo{journal}{Phys. Plasmas} \textbf{\bibinfo{volume}{9}},
  \bibinfo{pages}{4435} (\bibinfo{year}{2002}).

\bibitem[{\citenamefont{Wardle and Ng}(1999)}]{Wardle99}
\bibinfo{author}{\bibfnamefont{M.}~\bibnamefont{Wardle}} \bibnamefont{and}
  \bibinfo{author}{\bibfnamefont{C.}~\bibnamefont{Ng}},
  \bibinfo{journal}{Monthly Notices of the Royal Astronomical Society}
  \textbf{\bibinfo{volume}{303}}, \bibinfo{pages}{239} (\bibinfo{year}{1999}).

\bibitem[{\citenamefont{Balbus and Terquem}(2001)}]{Balbus01}
\bibinfo{author}{\bibfnamefont{S.}~\bibnamefont{Balbus}} \bibnamefont{and}
  \bibinfo{author}{\bibfnamefont{C.}~\bibnamefont{Terquem}},
  \bibinfo{journal}{Astrophys. J.} \textbf{\bibinfo{volume}{552}},
  \bibinfo{pages}{235} (\bibinfo{year}{2001}).

\bibitem[{\citenamefont{Sano and Stone}(2002{\natexlab{a}})}]{Sano02}
\bibinfo{author}{\bibfnamefont{T.}~\bibnamefont{Sano}} \bibnamefont{and}
  \bibinfo{author}{\bibfnamefont{J.~M.} \bibnamefont{Stone}},
  \bibinfo{journal}{Astrophys. J.} \textbf{\bibinfo{volume}{570}},
  \bibinfo{pages}{314} (\bibinfo{year}{2002}{\natexlab{a}}).

\bibitem[{\citenamefont{Sano and Stone}(2002{\natexlab{b}})}]{Sano02b}
\bibinfo{author}{\bibfnamefont{T.}~\bibnamefont{Sano}} \bibnamefont{and}
  \bibinfo{author}{\bibfnamefont{J.~M.} \bibnamefont{Stone}},
  \bibinfo{journal}{Astrophys. J.} \textbf{\bibinfo{volume}{577}},
  \bibinfo{pages}{534} (\bibinfo{year}{2002}{\natexlab{b}}).

\bibitem[{\citenamefont{Rudakov}(2001)}]{Rudakov01}
\bibinfo{author}{\bibfnamefont{L.~I.} \bibnamefont{Rudakov}},
  \bibinfo{journal}{Phys. Scripta} \textbf{\bibinfo{volume}{2001}},
  \bibinfo{pages}{158} (\bibinfo{year}{2001}).

\bibitem[{\citenamefont{Salimullah and Shukla}(1999)}]{Salimullah99}
\bibinfo{author}{\bibfnamefont{M.}~\bibnamefont{Salimullah}} \bibnamefont{and}
  \bibinfo{author}{\bibfnamefont{P.~K.} \bibnamefont{Shukla}},
  \bibinfo{journal}{Physics of Plasmas} \textbf{\bibinfo{volume}{6}},
  \bibinfo{pages}{686} (\bibinfo{year}{1999}).

\bibitem[{\citenamefont{Rosenberg and Shukla}(2000)}]{Rosenberg00}
\bibinfo{author}{\bibfnamefont{M.}~\bibnamefont{Rosenberg}} \bibnamefont{and}
  \bibinfo{author}{\bibfnamefont{P.~K.} \bibnamefont{Shukla}},
  \bibinfo{journal}{Journal of Geophysical Research}
  \textbf{\bibinfo{volume}{105}}, \bibinfo{pages}{23135}
  (\bibinfo{year}{2000}).

\bibitem[{\citenamefont{Rudakov}(2002)}]{Rudakov02}
\bibinfo{author}{\bibfnamefont{L.~I.} \bibnamefont{Rudakov}},
  \bibinfo{journal}{Phys. Scripta} \textbf{\bibinfo{volume}{2002}},
  \bibinfo{pages}{58} (\bibinfo{year}{2002}).

\bibitem[{\citenamefont{Newman et~al.}(1992)\citenamefont{Newman, Newman, and
  Lovelace}}]{Newman92}
\bibinfo{author}{\bibfnamefont{W.~I.} \bibnamefont{Newman}},
  \bibinfo{author}{\bibfnamefont{A.~L.} \bibnamefont{Newman}},
  \bibnamefont{and} \bibinfo{author}{\bibfnamefont{R.~V.~E.}
  \bibnamefont{Lovelace}}, \bibinfo{journal}{Astrophys. J.}
  \textbf{\bibinfo{volume}{392}}, \bibinfo{pages}{622} (\bibinfo{year}{1992}).

\bibitem[{\citenamefont{Harquist et~al.}(1997)\citenamefont{Harquist, Pilipp,
  and Havnes}}]{Harquist97}
\bibinfo{author}{\bibfnamefont{T.~W.} \bibnamefont{Harquist}},
  \bibinfo{author}{\bibfnamefont{W.}~\bibnamefont{Pilipp}}, \bibnamefont{and}
  \bibinfo{author}{\bibfnamefont{O.}~\bibnamefont{Havnes}},
  \bibinfo{journal}{Astrophysics and Space Science}
  \textbf{\bibinfo{volume}{246}}, \bibinfo{pages}{243} (\bibinfo{year}{1997}).

\bibitem[{\citenamefont{Draine et~al.}(1983)\citenamefont{Draine, Roberge, and
  Dalgarno}}]{Draine83}
\bibinfo{author}{\bibfnamefont{B.~T.} \bibnamefont{Draine}},
  \bibinfo{author}{\bibfnamefont{W.~G.} \bibnamefont{Roberge}},
  \bibnamefont{and} \bibinfo{author}{\bibfnamefont{A.}~\bibnamefont{Dalgarno}},
  \bibinfo{journal}{Astrophys. J.} \textbf{\bibinfo{volume}{264}},
  \bibinfo{pages}{485} (\bibinfo{year}{1983}).

\bibitem[{\citenamefont{Chapman}(1956)}]{Chapman56}
\bibinfo{author}{\bibfnamefont{S.}~\bibnamefont{Chapman}},
  \bibinfo{journal}{Nuovo Cimento} \textbf{\bibinfo{volume}{5}},
  \bibinfo{pages}{1385} (\bibinfo{year}{1956}).

\bibitem[{\citenamefont{Fr{\'{o}}mang et~al.}(2002)\citenamefont{Fr{\'{o}}mang,
  Terquem, and Balbus}}]{Fromang02}
\bibinfo{author}{\bibfnamefont{S.}~\bibnamefont{Fr{\'{o}}mang}},
  \bibinfo{author}{\bibfnamefont{C.}~\bibnamefont{Terquem}}, \bibnamefont{and}
  \bibinfo{author}{\bibfnamefont{S.~A.} \bibnamefont{Balbus}},
  \bibinfo{journal}{Monthly Notices of the Royal Astronomical Society}
  \textbf{\bibinfo{volume}{329}}, \bibinfo{pages}{18} (\bibinfo{year}{2002}).

\bibitem[{\citenamefont{Myers and Goodman}(1988)}]{Myers88}
\bibinfo{author}{\bibfnamefont{P.~C.} \bibnamefont{Myers}} \bibnamefont{and}
  \bibinfo{author}{\bibfnamefont{A.~A.} \bibnamefont{Goodman}},
  \bibinfo{journal}{Astrophys. J.} \textbf{\bibinfo{volume}{326}},
  \bibinfo{pages}{L27} (\bibinfo{year}{1988}).

\bibitem[{\citenamefont{Oppenheimer and Dalgarno}(1974)}]{Oppenheimer74}
\bibinfo{author}{\bibfnamefont{M.}~\bibnamefont{Oppenheimer}} \bibnamefont{and}
  \bibinfo{author}{\bibfnamefont{A.}~\bibnamefont{Dalgarno}},
  \bibinfo{journal}{Astrophys. J.} \textbf{\bibinfo{volume}{192}},
  \bibinfo{pages}{29} (\bibinfo{year}{1974}).

\bibitem[{\citenamefont{Spitzer and Tomasko}(1968)}]{Spitzer68}
\bibinfo{author}{\bibfnamefont{L.}~\bibnamefont{Spitzer}} \bibnamefont{and}
  \bibinfo{author}{\bibfnamefont{M.~G.} \bibnamefont{Tomasko}},
  \bibinfo{journal}{Astrophys. J.} \textbf{\bibinfo{volume}{152}},
  \bibinfo{pages}{971} (\bibinfo{year}{1968}).

\bibitem[{\citenamefont{Umebayashi and Nakano}(1990)}]{Umebayashi90}
\bibinfo{author}{\bibfnamefont{T.}~\bibnamefont{Umebayashi}} \bibnamefont{and}
  \bibinfo{author}{\bibfnamefont{T.}~\bibnamefont{Nakano}},
  \bibinfo{journal}{Monthly Notices of the Royal Astronomical Society}
  \textbf{\bibinfo{volume}{243}}, \bibinfo{pages}{236} (\bibinfo{year}{1990}).

\bibitem[{\citenamefont{Zweibel}(1999)}]{Zweibel99}
\bibinfo{author}{\bibfnamefont{E.~G.} \bibnamefont{Zweibel}},
  \bibinfo{journal}{Phys. Plasmas} \textbf{\bibinfo{volume}{6}},
  \bibinfo{pages}{1725} (\bibinfo{year}{1999}).

\bibitem[{\citenamefont{Mathis et~al.}(1977)\citenamefont{Mathis, Rumpl, and
  Nordsieck}}]{Mathis77}
\bibinfo{author}{\bibfnamefont{J.~S.} \bibnamefont{Mathis}},
  \bibinfo{author}{\bibfnamefont{W.}~\bibnamefont{Rumpl}}, \bibnamefont{and}
  \bibinfo{author}{\bibfnamefont{K.~H.} \bibnamefont{Nordsieck}},
  \bibinfo{journal}{Astrophys. J.} \textbf{\bibinfo{volume}{217}},
  \bibinfo{pages}{425} (\bibinfo{year}{1977}).

\bibitem[{\citenamefont{Chandrasekhar and Fermi}(1953)}]{ChandrasekharFermi53}
\bibinfo{author}{\bibfnamefont{S.}~\bibnamefont{Chandrasekhar}}
  \bibnamefont{and} \bibinfo{author}{\bibfnamefont{E.}~\bibnamefont{Fermi}},
  \bibinfo{journal}{Astrophys. J.} \textbf{\bibinfo{volume}{118}},
  \bibinfo{pages}{116C} (\bibinfo{year}{1953}).

\bibitem[{\citenamefont{Vainshtein et~al.}(2000)\citenamefont{Vainshtein,
  Chitre, and Olinto}}]{Vainshtein00}
\bibinfo{author}{\bibfnamefont{S.~I.} \bibnamefont{Vainshtein}},
  \bibinfo{author}{\bibfnamefont{S.~M.} \bibnamefont{Chitre}},
  \bibnamefont{and} \bibinfo{author}{\bibfnamefont{A.~V.}
  \bibnamefont{Olinto}}, \bibinfo{journal}{Phys. Rev. E}
  \textbf{\bibinfo{volume}{61}}, \bibinfo{pages}{4422} (\bibinfo{year}{2000}).

\bibitem[{\citenamefont{Desyatnikov et~al.}(2000)\citenamefont{Desyatnikov,
  Maimistov, and Malomed}}]{Desyatnikov00}
\bibinfo{author}{\bibfnamefont{A.}~\bibnamefont{Desyatnikov}},
  \bibinfo{author}{\bibfnamefont{A.}~\bibnamefont{Maimistov}},
  \bibnamefont{and} \bibinfo{author}{\bibfnamefont{B.}~\bibnamefont{Malomed}},
  \bibinfo{journal}{Phys. Rev. E} \textbf{\bibinfo{volume}{61}},
  \bibinfo{pages}{3107} (\bibinfo{year}{2000}).

\bibitem[{\citenamefont{Yakimenko et~al.}(2005)\citenamefont{Yakimenko,
  Zaliznyak, and Kivshar}}]{Yakimenko05}
\bibinfo{author}{\bibfnamefont{A.}~\bibnamefont{Yakimenko}},
  \bibinfo{author}{\bibfnamefont{Y.}~\bibnamefont{Zaliznyak}},
  \bibnamefont{and} \bibinfo{author}{\bibfnamefont{Y.}~\bibnamefont{Kivshar}},
  \bibinfo{journal}{Phys. Rev. E} \textbf{\bibinfo{volume}{71}}
  (\bibinfo{year}{2005}).

\bibitem[{\citenamefont{Lada and Fich}(1996)}]{Lada96}
\bibinfo{author}{\bibfnamefont{C.~J.} \bibnamefont{Lada}} \bibnamefont{and}
  \bibinfo{author}{\bibfnamefont{M.}~\bibnamefont{Fich}},
  \bibinfo{journal}{Astrophys. J.} \textbf{\bibinfo{volume}{459}},
  \bibinfo{pages}{638} (\bibinfo{year}{1996}).

\bibitem[{\citenamefont{Tamura et~al.}(1999)\citenamefont{Tamura, Hough,
  Greaves, Morino, Chrisostomou, Holland, and Momose}}]{Tamura99}
\bibinfo{author}{\bibfnamefont{M.}~\bibnamefont{Tamura}},
  \bibinfo{author}{\bibfnamefont{G.~H.} \bibnamefont{Hough}},
  \bibinfo{author}{\bibfnamefont{J.~S.} \bibnamefont{Greaves}},
  \bibinfo{author}{\bibfnamefont{J.-I.} \bibnamefont{Morino}},
  \bibinfo{author}{\bibfnamefont{A.}~\bibnamefont{Chrisostomou}},
  \bibinfo{author}{\bibfnamefont{W.~S.} \bibnamefont{Holland}},
  \bibnamefont{and} \bibinfo{author}{\bibfnamefont{M.}~\bibnamefont{Momose}},
  \bibinfo{journal}{Astrophys. J.} \textbf{\bibinfo{volume}{525}},
  \bibinfo{pages}{832} (\bibinfo{year}{1999}).

\bibitem[{\citenamefont{Lovelace et~al.}(1992)\citenamefont{Lovelace, Romanova,
  and Bisnovatyi-Kogan}}]{Lovelace92}
\bibinfo{author}{\bibfnamefont{R.~V.~E.} \bibnamefont{Lovelace}},
  \bibinfo{author}{\bibfnamefont{M.~M.} \bibnamefont{Romanova}},
  \bibnamefont{and} \bibinfo{author}{\bibfnamefont{G.~S.}
  \bibnamefont{Bisnovatyi-Kogan}}, \bibinfo{journal}{Monthly Notices of the
  Royal Astronomical Society} \textbf{\bibinfo{volume}{275}},
  \bibinfo{pages}{244} (\bibinfo{year}{1992}).

\bibitem[{\citenamefont{Ustyugova et~al.}(2000)\citenamefont{Ustyugova,
  Lovelace, Romanova, Li, and Colgate}}]{Ustyugova00}
\bibinfo{author}{\bibfnamefont{G.~V.} \bibnamefont{Ustyugova}},
  \bibinfo{author}{\bibfnamefont{R.~V.~E.} \bibnamefont{Lovelace}},
  \bibinfo{author}{\bibfnamefont{M.~M.} \bibnamefont{Romanova}},
  \bibinfo{author}{\bibfnamefont{H.}~\bibnamefont{Li}}, \bibnamefont{and}
  \bibinfo{author}{\bibfnamefont{S.~A.} \bibnamefont{Colgate}},
  \bibinfo{journal}{Astrophys. J.} \textbf{\bibinfo{volume}{541}},
  \bibinfo{pages}{L21} (\bibinfo{year}{2000}).

\bibitem[{\citenamefont{Ciardi et~al.}(2006)\citenamefont{Ciardi, Lebedev,
  Stehle, and Lery}}]{Ciardi06}
\bibinfo{author}{\bibfnamefont{A.}~\bibnamefont{Ciardi}},
  \bibinfo{author}{\bibfnamefont{S.~V.} \bibnamefont{Lebedev}},
  \bibinfo{author}{\bibfnamefont{C.}~\bibnamefont{Stehle}}, \bibnamefont{and}
  \bibinfo{author}{\bibfnamefont{T.}~\bibnamefont{Lery}}, in
  \emph{\bibinfo{booktitle}{Journal de Physique IV (Proceedings)}}
  (\bibinfo{organization}{LUTH, Observatoire de Paris, UMR 8102 CNRS, 92195
  Meudon, France; The Blackett Laboratory, Imperial College, London SW7 2BW,
  UK}, \bibinfo{year}{2006}), pp. \bibinfo{pages}{1043--1045}.

\bibitem[{\citenamefont{Ciardi et~al.}(2009)\citenamefont{Ciardi, Lebedev,
  Frank, Suzuki-Vidal, Hall, Bland, Harvey-Thompson, Blackman, and
  Camenzind}}]{Ciardi09}
\bibinfo{author}{\bibfnamefont{A.}~\bibnamefont{Ciardi}},
  \bibinfo{author}{\bibfnamefont{S.~V.} \bibnamefont{Lebedev}},
  \bibinfo{author}{\bibfnamefont{A.}~\bibnamefont{Frank}},
  \bibinfo{author}{\bibfnamefont{F.}~\bibnamefont{Suzuki-Vidal}},
  \bibinfo{author}{\bibfnamefont{G.~N.} \bibnamefont{Hall}},
  \bibinfo{author}{\bibfnamefont{S.~N.} \bibnamefont{Bland}},
  \bibinfo{author}{\bibfnamefont{A.}~\bibnamefont{Harvey-Thompson}},
  \bibinfo{author}{\bibfnamefont{E.~G.} \bibnamefont{Blackman}},
  \bibnamefont{and}
  \bibinfo{author}{\bibfnamefont{M.}~\bibnamefont{Camenzind}},
  \bibinfo{journal}{Astrophys. J.} \textbf{\bibinfo{volume}{691}},
  \bibinfo{pages}{L147} (\bibinfo{year}{2009}).

\end{thebibliography}

\appendix*
\section{Variational Method For Trial Functions of GGS} \label{sec:variations}
Here, we demonstrate localized 3D soliton solutions of the GGS equation with nonlinear current term, with the cylindrical coordinate and spherical coordinate trial functions. This variational method and trial functions were borrowed from \cite{Desyatnikov00}. Furthermore, the integration of the one-dimensional functions was performed using a fourth-order Runge-Kutta method from $r = 0$ to $r = 10$ for the cylindrical ansatz and $\rho = 0$ to $\rho = 10$ for the spherical ansatz.

\subsection{Cylindrical Coordinate Ansatz}
The Lagrangian of the differential expression (\ref{eq:soliton0}) is given by,
\begin{eqnarray}
    &&\begin{aligned}{}
    L\paren{\partial_r A_0, \partial_z A_0, A_0; r, z} = \frac{1}{2}r\paren{\partial_r A_0}^2 + \frac{1}{2}r\paren{
    \partial_z A_0}^2 + \frac{A_0^2}{2r} + \frac{1}{2}r^3 A_0^2 -
    \frac{1}{2}r^3 A_0^4 \end{aligned} \label{eq:cylinLag}
\end{eqnarray}
With this ansatz,
\begin{eqnarray}
    &&A_0(r,z) = U(r)\sech\paren{\mu z}, \label{eq:ansatz1}
\end{eqnarray}
we get the averaged Lagrangian ${\mathcal L}_1$, where $U' = dU/dr$:
\begin{eqnarray}
    &&\begin{aligned}{}
    {\mathcal L}_1 = \int_{-\infty}^{\infty} 
    L\paren{\partial_r A_0, \partial_z A_0, A_0; r, z}\,dz = \frac{r}{\mu}\paren{\di{U}{r}}^2 + \frac{\mu r}{3}U^2 + 
    \frac{U^2}{\mu r} + \frac{r^3}{\mu}U^2 - \frac{2r^3}{3\mu}U^4
    \end{aligned} \label{eq:avgLag1}
\end{eqnarray}
If we vary ${\mathcal L}$ with respect to $U$ and $dU/dr$ we then have a differential equation in $r$:
\begin{eqnarray}
    &&\din{U}{r}{2} + \frac{1}{r}\di{U}{r} - \frac{U}{r^2} -
    \frac{1}{3}\mu^2 U - r^2 U + \frac{4}{3}r^2 U^3 = 0
    \label{eq:difeq1}
\end{eqnarray}
The solutions that we look for are given by the
\begin{eqnarray}
    &&\begin{aligned}{}
    &\lim_{r\to 0} U(r) = U'(0)r \\
    &\lim_{r\to\infty} U(r) = r^{-1} \exp\paren{-\frac{1}{2}r^2} 
    \end{aligned} \label{eq:asymptot1}
\end{eqnarray}
We employ the shooting method, with free parameter $U'(0)$, by
integrating out Eq.~(\ref{eq:difeq1}) from $r = 0$ to some sufficienly
large $r = r_{\text{lim}} \gg w$, where $w$ is the width of the
structure. Furthermore, since we postulate the lowest-energy solutions
are probably dominant, we look for solutions that have a single node
in $U(r)$.

For each $\mu$ there is a unique single-node solution $U(r;\mu)$
satisfying Eq.~(\ref{eq:difeq1}). To solve for this problem, we
require that the averaged Lagrangian Eq.~(\ref{eq:avgLag1}), averaged
over $r$, must be extremized with respect to $\mu$:
\begin{eqnarray}
    &&\int_0^{\infty} \parti{{\mathcal L}_1}{\mu}\,dr = 0
\end{eqnarray}
If we use these functions:
\begin{eqnarray}
    &&\begin{aligned}{} &\epsilon_1\paren{\mu} = \int_0^{\infty}
    rU\paren{r;\mu}^2\,dr \\ &\epsilon_2\paren{\mu} = \int_0^{\infty}
    r^3U\paren{r;\mu}^2\,dr \\ &\epsilon_3\paren{\mu} =
    \int_0^{\infty} r^3 U\paren{r;\mu}^4\,dr
    \end{aligned} \label{eq:epsilons}
\end{eqnarray}
We then get the relation,
\begin{eqnarray}
    &&\begin{aligned}{} &\int_0^{\infty} r\paren{\paren{\di{U}{r}}^2 +
    \frac{U^2}{r^2}}\,dr = \frac{1}{3}\mu^2\epsilon_1\paren{\mu} -
    \epsilon_2\paren{\mu} + \frac{2}{3}\epsilon_3\paren{\mu}.
    \end{aligned} \label{eq:intermediate1step1}
\end{eqnarray}
However, using (\ref{eq:difeq1}) one can derive this relation,
\begin{eqnarray}
    &&\begin{aligned}{}
    r\brak{\paren{\di{U}{r}}^2 + \frac{U^2}{r^2}} = \di{}{r}\paren{rU\di{U}{r}} - \frac{1}{3}\mu^2 rU^2 - r^3 U^2 +
    \frac{4}{3}r^3 U^4. \end{aligned}\label{eq:intermediate1step2}
\end{eqnarray}
Substituting Eq.~(\ref{eq:intermediate1step2}) into the integral on the left hand side of Eq.~(\ref{eq:intermediate1step1}), and noting that the first right hand side term in Eq.~(\ref{eq:intermediate1step2}) integrates to zero because it is a surface term, we then get the residual function for $\mu_0$:
\begin{eqnarray}
    &&\delta_1\paren{\mu_0} = \mu_0^2\epsilon_1\paren{\mu_0} - \epsilon_3\paren{\mu_0} = 0 \label{eq:delta1}
\end{eqnarray}
Our estimates of $\mu_0$ (to four decimal places) and $U_0'$ that satify (\ref{eq:delta1}) are given below:
\begin{eqnarray}
    &&\begin{aligned}{}
    \mu_0 &\approx 2.1852 \\
    U'(0) &\approx 1.0115
    \end{aligned} \label{eq:approx1}
\end{eqnarray}

\subsection{Spherical Coordinate Ansatz}
For the spherical coordinate ansatz, we transform (\ref{eq:soliton0}) into spherical coordinates, with $\rho = \sqrt{r^2 + z^2}$ and $\cos\theta = z/\sqrt{r^2 + z^2}$:
\begin{eqnarray}
    &&\begin{aligned}{}
    &\frac{1}{\rho^2}\parti{}{\rho}\paren{\rho^2\parti{A_0}{\rho}} + 
    \frac{1}{\rho^2\sin\theta}\parti{}{\theta}\paren{\sin\theta
    \parti{A_0}{\theta}} - \\
    &\frac{A_0}{\rho^2\sin^2\theta} -
    \rho^2\sin^2\theta A_0 + 
    2\rho^2\sin^2\theta A_0^3 = 0
    \end{aligned} \label{eq:spherDE}
\end{eqnarray}
Which has the Lagrangian,
\begin{eqnarray}
    &&\begin{aligned}{}
    &L\paren{\partial_{\rho} A_0, \partial_{\theta} A_0, A_0;\rho,\theta} = 
    \frac{1}{2}\rho^2\sin\theta\paren{\parti{A_0}{\rho}}^2 + \\
    &\frac{1}{2}\sin\theta\paren{\parti{A_0}{\theta}}^2 + 
    \frac{A_0^2}{2\sin\theta} + \frac{1}{2}\rho^4\sin^3\theta A_0^2 - \frac{1}{2}\rho^4\sin^3\theta A_0^4.
    \end{aligned} \label{eq:spherLag}
\end{eqnarray}
With the ansatz:
\begin{eqnarray}
    &&A_0(\rho,\theta) = U(\rho)\sin\theta \label{eq:ansatz3}
\end{eqnarray}
We get the averaged Lagrangian:
\begin{eqnarray}
    &&\begin{aligned}{}
    {\mathcal L}_2 = \int_0^{\pi} 
    L\paren{\partial_{\rho} A_0, \partial_{\theta} A_0,
    A_0;\rho,\theta} \,d\theta = \frac{2}{3}\rho^2\paren{\di{U}{\rho}}^2 + \frac{4}{3}U^2
    + \frac{8}{15}\rho^4 U^2 - \frac{16}{35}\rho^4 U^4
    \end{aligned} \label{eq:avgLag3}
\end{eqnarray}
That results in this differential expression:
\begin{eqnarray}
    &&\din{U}{\rho}{2} + \frac{2}{\rho}\di{U}{\rho} - 
    \frac{2 U}{\rho^2} - \frac{4}{5}\rho^2 U + 
    \frac{48}{35}\rho^2 U^3 = 0 \label{eq:difeq3}
\end{eqnarray}
We shoot with $U(0) = 0$ and free parameter $U'(0)$ such that for sufficiently large $\rho$, we get $U(\rho) \to \alpha\rho^{-1}\exp\paren{-\rho^2/5}$, where $\alpha$ is some constant. $U'(0)$ is given by,
\begin{eqnarray}
    &&U'(0) \approx 1.5897 \label{eq:param3}
\end{eqnarray}
\end{document}